# Generation and manipulation of nonclassical light using photonic crystals

Jelena Vuckovic[a,1], Dirk Englund[a], David Fattal[b], Edo Waks[a], Yoshihisa Yamamoto[b]

[a] *Ginzton Laboratory, Stanford University, Stanford, CA 94305-4088, USA*

[b] *Quantum Entanglement Project, Ginzton Laboratory, Stanford University, Stanford, CA 94305-4088, USA*

**Abstract**

Photonic crystal cavities can localize light into nanoscale volumes with high quality factors. This permits a strong interaction between light and matter, which is important for the construction of classical light sources with improved properties (e.g., low threshold lasers) and of nonclassical light sources (such as single and entangled photon sources) that are crucial pieces of hardware of quantum information processing systems. This article will review some of our recent experimental and theoretical results on the interaction between single quantum dots and photonic crystal cavity fields, and on the integration of multiple photonic crystal devices into functional circuits for quantum information processing.

## 1. Introduction

Photonic crystal (PC) cavities enable localization of light into mode volumes (V) below a cubic optical wavelength (smaller than any other types of optical resonators) with high quality factors (Q) [1-3]. A strong localization of light in PC cavities results from Distributed Bragg Reflection (DBR) as the principal confinement mechanism, as opposed to, e.g., microspheres or microdisks, which employ Total Internal Reflection (TIR). How good a resonator is for a particular application depends on the ratio of powers of Q and V. For example, for spontaneous emission rate enhancement through the Purcell effect, one desires maximal Q/V; for nonlinear optical effects $Q^2/V$; while for the strong coupling regime of cavity QED, maximizing ratios $g/\kappa \sim Q/\sqrt{V}$ and $g/\gamma \sim 1/\sqrt{V}$ is important. In these expressions, V is the cavity mode volume:

$$V = \left[\int \varepsilon(\vec{r})|\vec{E}(\vec{r})|^2 d^3\vec{r}\right] / \max\left\{\varepsilon(\vec{r})|\vec{E}(\vec{r})|^2\right\} \quad (1)$$

g is the emitter-cavity field coupling, and $\kappa$ and $\gamma$ are the cavity field and emitter dipole decay rates, respectively [1].

One of many devices whose quality is improved by application of nanocavities is a single-photon source. Single photons on demand can be generated by combining pulsed excitation of a single quantum emitter (such as a quantum dot (QD), an atom, a molecule, or a nitrogen vacancy center in diamond) and spectral filtering [4-7]. When a QD is excited with a short (fs or ps) laser pulse, electron-hole pairs are created within it; the created carriers recombine in a radiative cascade, leading to the generation of several photons for each laser pulse. All of these photons have slightly different frequencies, resulting from discrete energy states of the low-dimensional system and the Coulomb interaction among carriers. The last emitted photon for each pulse has a unique frequency, and can be spectrally isolated. In the scheme we employ, the excitation laser pulse frequency can be either resonant with a higher-order transition of a QD or above the band gap of the material surrounding it (this technique is also referred to as incoherent excitation). Three criteria are taken into account when evaluating the quality of a single-photon source: efficiency, photon statistics (i.e., multi-photon probability suppression measured by the second-order coherence function $g^2(0)$), and quantum indistinguishability (i.e., the mean wavefunction overlap between emitted photons). The requirements depend on the specific application; for example, $g^2(0)$ should be as small as possible for quantum key distribution [8], efficiency should be high, but indistinguishability is not as important. However, for photonic quantum computation, all of these criteria are critical [9]. While multi-photon probability suppression ($g^2(0)$) is already small for a single quantum emitter excited using resonant-excitation methods [4], single photon efficiency and indistinguishability are poor, as photons are emitted in random directions in space and dephasing mechanisms are strong. However, both efficiency and indistinguishability can be improved by embedding a quantum emitter into a cavity (with high quality factor Q and small mode volume V), where the spontaneous emission rate of the emitter can be enhanced relative to its value in bulk (or free-space) as a result of its coupling to the cavity mode (Purcell effect). The spontaneous emission rate of an emitter into the cavity mode, $\Gamma_{cav}$, is enhanced relative to its value without a cavity $\Gamma_0 = \omega^3 \mu^2 n / (3\pi\varepsilon_0 \hbar c^3)$ by the Purcell factor:

$$\frac{\Gamma_{cav}}{\Gamma_0} = \frac{3\lambda_c^3}{4\pi^2 n^3} \cdot \frac{Q}{V_0} \cdot \left|\frac{\vec{E} \cdot \vec{\mu}}{E_{max}\mu}\right|^2 \frac{\Delta\lambda_c^2}{\Delta\lambda_c^2 + 4(\lambda - \lambda_c)^2} \quad (2)$$

---

[1] Corresponding author. E-mail: jela@stanford.edu, fax: 1-650-723-5320

The Purcell factor is maximized for an emitter on resonance with the cavity mode spectrally ($\lambda=\lambda_C$) and spatially (i.e., when it is located at the cavity field maximum, where $E=E_{max}$), and with dipole moment aligned with the electric field ($\vec{\mu} \parallel \vec{E}$). The total spontaneous emission rate $\Gamma=\Gamma_{cav}+\Gamma_{other}$ is then primarily determined by $\Gamma_{cav}$, since the spontaneous emission rate into all other modes is small: $\Gamma_{other}<<\Gamma_{cav}$. The fraction of the spontaneously emitted photons that are coupled into a single cavity mode is given by $\beta=\Gamma_{cav}/(\Gamma_{cav}+\Gamma_{other})$, which also increases with $\Gamma_{cav}$. The external efficiency of the single photon source is then $\eta=\beta\eta_{extract}$, where $\eta_{extract}$ is the extraction efficiency, i.e., the fraction of photons coupled to the cavity mode that are redirected towards a particular output where they can be collected. In addition, as a result of the Purcell effect, the radiative lifetime is reduced significantly below the dephasing time, leading to an increase in the indistinguishability of emitted photons (and also an increase in the possible repetition rate of the source). In case of the incoherent excitation described above, the indistinguishability is given by [7]:

$$I = \frac{\Gamma}{\Gamma + \alpha} \frac{\delta}{2\Gamma + \delta} \quad (3)$$

where $\alpha$ is the dephasing rate of the excited state, and $\delta$ is the relaxation rate from the higher-order excited state to the first excited state (from which the single-photon pulse is emitted), leading to a jitter in the arrival time of the single photon wavepacket. Therefore, in order to maximize $I$, the radiative lifetime ($1/\Gamma$) should be reduced below $1/\alpha$ (0.5-1ns), but should be well above $1/\delta$ (~10ps) [7]. To satisfy the first condition, the Purcell effect has to be employed, since an InAs QD in bulk GaAs has radiative lifetime between 1-2ns. However, this would lead to an improvement in $I$ only when the radiative lifetime is well above $1/\delta$, as determined by the 2nd part of the expression. By differentiating the expression for $I$ with respect to $\Gamma$, we can conclude that $I$ can be maximized to around 70-80% in case of an incoherent excitation combined with Purcell effect. This is achieved when the radiative lifetime is reduced to about 100-140ps, i.e., $\Gamma = \sqrt{\alpha\delta/2}$, in which case $I = 1/\left(1+\sqrt{2\alpha/\delta}\right)^2$.

## 2. Quantum dot –photonic crystal cavity single photon source

By employing quantum dots inside micropost microcavities, we were able to significantly improve the properties of a single photon source [5-7]: the multiphoton probability suppression relative to a Poisson-distributed source (attenuated laser) of the same intensity was as small as $g^2(0)\approx 2\%$, the duration of single photon pulses was reduced to ≈200 ps, the spontaneous emission coupling factor and external efficiency improved to $\beta\approx 85\%$ and $\eta\approx 35\%$, and the measured indistinguishability between two consecutively emitted photons I≈81%. Further improvement in efficiency and indistinguishability of a single photon source can be achieved by employing better microcavities, with larger Q/V ratios, and consequently stronger emitter-cavity coupling. For example, photonic crystal microcavities shown in Fig.1 have an order of magnitude higher Q/V ratios than the best microposts [10].

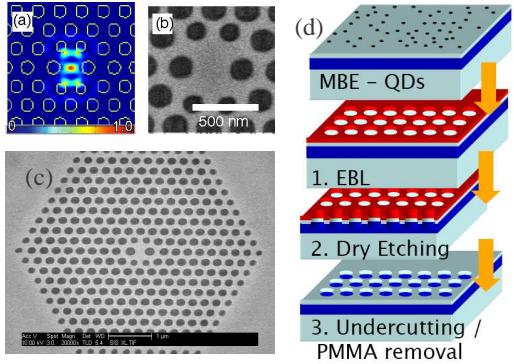

**Fig.1. (a-c)** E-field pattern and SEM micrographs of fabricated photonic crystal cavities such as the ones used in the experiments (scalebar in (c) is 1μm). Cavities support the dipole mode (a) whose mode volume is of the order of a cubic optical wavelength ($V\sim 1/2(\lambda/n)^3$). Structures are made of GaAs containing a single layer of InAs QDs embedded at the central plane of a PC membrane. **(d)** Fabrication procedure for photonic crystal microcavities used in the experiments. A 3-layer wafer (GaAs substrate, AlAs sacrificial layer, and ~165nm GaAs membrane containing InAs QDs at the central plane) is grown by MBE. Fabrication starts with spinning PMMA on the top surface and e-beam lithography of desired structures, and development of PMMA (1). Developed PMMA is used as a mask for dry-etching by ECR RIE (2). Finally, PMMA is removed by oxygen plasma, and wafers are undercut by removing the AlAs sacrificial layer in HF. This results in free-standing PC membranes.

The fabrication procedure for photonic crystal cavities and fabricated structures is shown in Fig.1. GaAs wafers (including a single InAs QD layer with density of ~100μm$^{-2}$) are usually grown by Molecular Beam Epitaxy (MBE), and PC cavities are fabricated by a combination of electron-beam lithography, dry- and wet-etching. Resulting structures have central defect regions of the size of 200-300nm, where the dipole mode is localized (mode volume $V\sim 1/2(\lambda/n)^3$). Such structures are placed in a He-flow cryostat, and excited in the direction perpendicular to the surface with a mode-locked Ti:sapphire laser with 160fs pulses, 13ns repetition period, and at the wavelength ~750nm, above the band-gap of GaAs. The emission is also collected in the vertical direction and sent towards the spectrometer/streak camera for time-resolved photoluminescence (PL) measurements, or Hanbury Brown and Twiss setup for measurements of the 2$^{nd}$ order correlation function $g^2(t)$ (for a detailed description of our experimental apparatus, please refer to [11]).

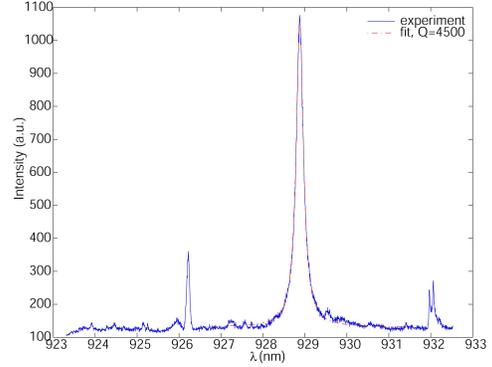

**Fig. 2.** Measured Q=4500 in a GaAs PC cavity with embedded InAs QDs (blue). The cavity supports the dipole mode (λ~929nm) and mode volume $V\sim 1/2(\lambda/n)^3$. The red curve is the Lorentzian fit to the measured cavity resonance.

Recently, we have also demonstrated single photon generation on demand from a QD embedded in such a PC microcavity, using incoherent excitation techniques [11]. Fig. 1 shows the photonic crystal cavities used in the experiment, and Fig. 2 shows the PL spectum of such a cavity at high excitation power indicating Q=4500 with the mode volume $V\approx 1/2(\lambda/n)^3$ (the dipole mode field pattern is shown in Fig. 1a). Fig. 3 shows the control of the radative lifetime of QDs embedded in such a cavity: in this case, for QD A coupled to the cavity resonance with Q=250 (see Fig. 4), the lifetime is reduced to 650ps, while for the QDs B and C which are spectrally detuned from the cavity resonance, lifetimes increase to 3.8ns and 4.2 ns, respectively. The suppression of the spontaneous emission is a result of the reduction in the photon density of states inside the photonic band gap [11].

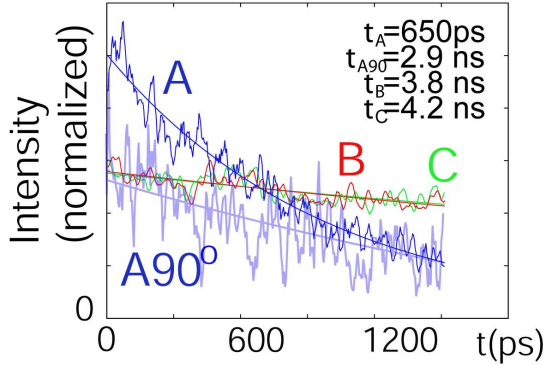

**Fig. 3.** Measured radiative lifetime reduction of individual QDs embedded in a PC cavity, such as the one from Fig. 1 (the lifetime is reduced relative to a bulk QD lifetime of ~1.7ns). For QD A resonant with the cavity with Q~250 lifetime is reduced to 650ps, while for the off-resonant QDs B and C lifetime increases to 3.8ns and 4.2ns, respectively (the spectroscopic measurements on the same QDs are shown in Fig. 4). In addition, the emission from QD A which is cross-polarized relative to the cavity mode is also suppressed, with lifetime of 2.9ns.

In the experiments shown in Fig. 3, we spectrally probe a single QD embedded inside the PC cavity. Therefore, by combining a pulsed excitation of such a QD and spectral filtering, we can generate single photons on demand, whose pulse duration is controlled between 200ps and 8ns by the Purcell effect. This is shown in Fig. 4, for the same QDs A and B.

With such a radiative lifetime reduction, the photon indistinguishability under incoherent excitation is still limited to 70-80%, as given by Eq. (3). One possible approach to increase the indistinguishability is to employ *coherent QD excitation techniques* (adiabatic Raman passage) borrowed from atomic physics [12].

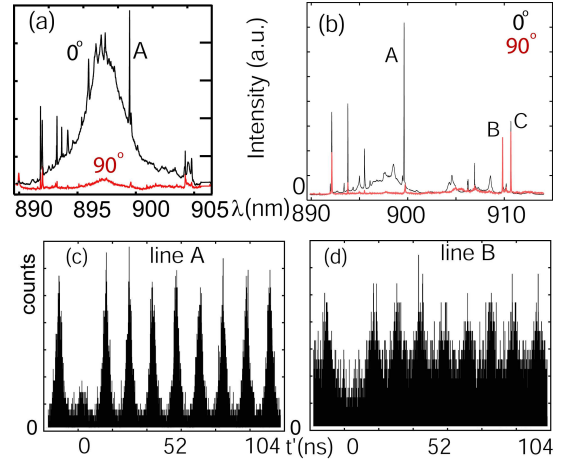

**Fig. 4.** (a) The high pump-intensity spectrum of a PC cavity reveals cavity resonance. The mode is the primarily linearly polarized dipole from Fig. 1, which is confirmed by the polarization measurement, and has Q~250. (b) The low-intensity spectrum shows the coupled QD line A and uncoupled lines B and C (small wavelength offset for clarity). (c),(d) $g^{(2)}(t)$ measurements for lines A and B, respectively, indicating the suppression of multiphoton probability, as indicated by the suppression of the central peak ($g^{(2)}(0)<0.5$). From the broadening of the side peaks in $g^{(2)}(t)$, we can also infer the lifetime of the QDs A and B, which is measured to be 650ps and 3.8ns, respectively (as shown in Fig. 3).

However, this requires a strong coupling regime of a single QD to the cavity field and a three-level system in the lambda configuration. A three-level system can be achieved by loading of an excess conduction band electron into a QD and applying a magnetic-field in the x-direction, which causes Zeeman splitting of the spin-up and –down levels in the conduction band [13,14]. The strong coupling regime of the cavity QED is achievable in PC cavities, such as the one from Fig. 1; the demonstrated set of parameters in our experiments (Q~5000, V~0.5$(\lambda/n)^3$) should be sufficient to observe strong coupling between a *single* QD on resonance spectrally and spatially with the cavity mode, since the emitter-cavity field coupling strength $g = \mu/\hbar \sqrt{\hbar\omega/2\varepsilon V} \approx 380 GHz$ is larger than the cavity field decay rate $\kappa=\omega/2Q\approx240$GHz and the excitonic dipole decay rate $\gamma$~2GHz [11]. Gated experiments on bulk QDs [15] suggest that it may be possible to bring a QD on resonance with the cavity mode via the DC stark shift. In addition, methods for spatial positioning of PC cavities to QDs have also been demonstrated recently [2].

From Eq. (3), it follows that another way to improve photon indistinguishability is to use incoherent excitation, but to increase the rate δ, which would in turn reduce the jitter at the beginning of the single photon pulse. This approach is more speculative, but would lead to a simpler QD excitation technique, and would eliminate the need for the magnetic field and precise single electron loading into a QD. An increase in $\delta$ can be achieved by introducing nanostructures to control the phonon density of states (such as the phonon cavity already demonstrated in [16]). In this case, an increase in the density of states of phonons mediating the transition from the higher order excited state of the QD to its first excited state would lead to an increase in δ, and a consequent improvement in $I$. If this transition is for example enhanced by a modest factor of 10, we could already reach $I = 1/\left(1+\sqrt{2\alpha/\delta}\right)^2 \approx 90\%$ with this simple excitation technique. In this case, the radiative lifetime would also have to be reduced to $1/\Gamma = \sqrt{2/\alpha\delta} \approx 40 ps$ by Purcell effect, which is very reasonable with our present quality of PC cavities [11].

## 3. Construction of high-efficiency single photon sources

As mentioned above, photonic crystal cavities also suppress the emission into other modes as a result of the photonic band gap ($\Gamma_{other}/\Gamma_0$=0.2), and enhance the emission into the cavity mode [11]. Therefore, for a moderate 8-fold spontaneous emission rate enhancement observed in the experiment, we can already reach $\beta=\Gamma_{cav}/(\Gamma_{cav}+\Gamma_{other})$=98%. However, preliminary measurements indicate that the total efficiency is only of the order of 20%, which is a result of the low $\eta_{extract}$. As shown in Fig. 5, in vertically symmetric structure the radiation pattern is symmetric in the direction perpendicular to the PC surface, implying that in the upward direction we can collect a maximum of 50% of the emitted photons (this is the direction in which we are probing PC cavities in our present experiments). There are two possible ways to improve this: employment of the DBR under the structure (see Fig. 5), or the lateral outcoupling strategy that we recently studied theoretically (Fig. 6) [17]. Our Finite-Difference Time-Domain simulations indicate that a GaAs/AlAs DBR placed underneath a PC membrane at about half-wavelength (see Fig. 5) can redirect more than 90% of the emitted photons upwards, without any effect on the cavity Q.

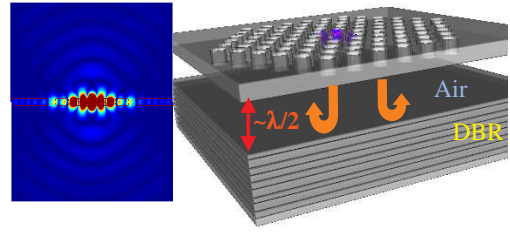

**Fig. 5 (Left)** Radiation pattern of a PC cavity such as the one from Fig. 1 in the direction perpendicular to PC membrane, revealing upward and downward radiation lobes. This limits the collection efficiency in the upward direction to 50% in the vertically symmetric structures **(Right)** Proposal for introducing asymmetry into the structure in the vertical direction, by placing a DBR underneath it at about half wavelength (DBR is grown between GaAs substrate and AlAs sacrificial layer in Fig. 1). By tuning the distance between the DBR and the PC membrane (i.e., a thickness of sacrificial layer), more than 90% of the photons can be outcoupled upwards, without an effect on the cavity Q.

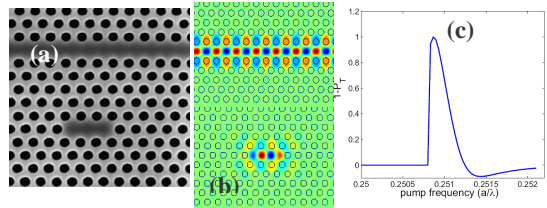

**Fig. 6 (a)** SEM image of the fabricated PC cavity, with lateral outcoupling into a PC waveguide (the material is GaAs containing InAs QDs). **(b)** $B_z$ field patterns at the slab center for both a PC-cavity and a waveguide. **(c)** Inverted normalized transmission through the PC waveguide indicating the dropping of the signal to the PC cavity side-coupled to it [17].

To achieve $\eta_{extract}$ close to 100%, we can also employ outcoupling from a cavity into a waveguide in the lateral direction [17]. An additional benefit of this approach is that photons outcoupled to waveguide can be easily redirected to other quantum gates on the same chip, thereby facilitating implementation of integrated quantum information processing systems. Our initial theoretical and fabrication results for the side-coupled cavity-waveguide configuration are shown in Fig. 6. The normalized PC waveguide transmission spectrum shown in Fig. 6c indicates the dropping of the signal to a PC cavity side-coupled to it, when they are on resonance (by optimization of the cavity and waveguide structures, even 100% dropping efficiency can be achieved).

## 4. Conclusion

Photonic crystal cavities with embedded quantum dots have shown an excellent potential for implementation of various devices for quantum information processing. For example, the spontaneous emission rate of a single quantum dot can be greatly modified, leading to increase in both indistinguishability and efficiency of single photons on demand emitted by a combination of QD pulsed excitation and spectral filtering. Lateral outcoupling strategies from a photonic crystal cavity into waveguide have additional advantages for implementation of quantum information processing systems on a chip, consisting of a combination of PC devices and QDs.

**Acknowledgment**

The authors would like to thank B. Zhang, G. Solomon (Stanford), T. Nakaoka and Y. Arakawa (U. of Tokyo) for the MBE/MOCVD growth of QD wafers. This work has been supported by the MURI Center for Photonic Quantum Information Systems (ARO/ARDA Program No. DAAD19-03-1-0199), NSF Grants ECS-0424080 and ECS-0421483, NDSEG and DCI fellowships.